\documentclass[sigconf]{acmart}
\AtBeginDocument{%
  \providecommand\BibTeX{{%
    \normalfont B\kern-0.5em{\scshape i\kern-0.25em b}\kern-0.8em\TeX}}}

\copyrightyear{2024}
\acmYear{2024}
\setcopyright{rightsretained}
\acmConference[CHI '24]{Proceedings of the CHI Conference on Human Factors in Computing Systems}{May 11--16, 2024}{Honolulu, HI, USA}
\acmBooktitle{Proceedings of the CHI Conference on Human Factors in Computing Systems (CHI '24), May 11--16, 2024, Honolulu, HI, USA}
\acmDOI{10.1145/3613904.3642260}
\acmISBN{979-8-4007-0330-0/24/05}
\acmPrice{15.00}
\acmISBN{978-1-4503-XXXX-X/18/06}

\usepackage{tabularray}
\usepackage[draft,inline,nomargin,index]{fixme}
\fxsetup{theme=colorsig,mode=multiuser,inlineface=\itshape,envface=\itshape}

\begin{document}

\title{``It doesn't tell me anything about how my data is used'': User Perceptions of Data Collection Purposes}



\author{Lin Kyi}
\affiliation{%
  \institution{Max Planck Institute for Security and Privacy}
  \city{Bochum}
  \country{Germany}}
\email{lin.kyi@mpi-sp.org}

\author{Abraham Mhaidli}
\affiliation{%
  \institution{University of Michigan}
  \city{Ann Arbor}
  \country{United States}}
\email{mhaidli@umich.edu}

\author{Cristiana Santos}
\affiliation{%
  \institution{Utrecht University}
  \city{Utrecht}
  \country{The Netherlands}}
\email{c.teixeirasantos@uu.nl}

\author{Franziska Roesner}
\affiliation{%
  \institution{University of Washington}
  \city{Seattle}
  \country{United States}}
\email{franzi@cs.washington.edu}

\author{Asia Biega}
\affiliation{%
  \institution{Max Planck Institute for Security and Privacy}
  \city{Bochum}
  \country{Germany}}
\email{asia.biega@mpi-sp.org}

\renewcommand{\shortauthors}{Kyi et al.}

\begin{abstract}
Data collection purposes and their descriptions are presented on almost all privacy notices under the GDPR, yet there is a lack of research focusing on how effective they are at informing users about data practices. We fill this gap by investigating users' perceptions of data collection purposes and their descriptions, a crucial aspect of informed consent. We conducted 23 semi-structured interviews with European users to investigate user perceptions of six common purposes (\textit{Strictly Necessary, Statistics and Analytics, Performance and Functionality, Marketing and Advertising, Personalized Advertising,} and \textit{Personalized Content}) and identified elements of an effective purpose name and description.

We found that most purpose descriptions do not contain the information users wish to know, and that participants preferred some purpose names over others due to their perceived transparency or ease of understanding. Based on these findings, we suggest how the framing of purposes can be improved toward meaningful informed consent. 
\end{abstract}

\ccsdesc[500]{Human-centered computing~Empirical studies in HCI}
\ccsdesc[500]{Security and privacy~Usability in security and privacy}
\ccsdesc[500]{Applied computing~Law}

\keywords{GDPR, personal data, purposes, privacy, tracking, qualitative methods}


\received{20 February 2007}
\received[revised]{12 March 2009}
\received[accepted]{5 June 2009}

\maketitle

\section{Introduction}

With the enforcement of the European Union's General Data Protection Regulation (GDPR)~\cite{EUdataregulations2018} in 2018 and the ePrivacy Directive~\cite{ePD}, privacy notices (also known as cookie banners or consent notices) have become the de facto standard for informing and collecting consent from EU and UK users. This has created a new industry of GDPR compliance tools, such as IAB Europe's Transparency and Consent Framework (TCF) and Consent Management Platforms (CMPs), established to help organizations manage their GDPR and ePD compliance through privacy notices~\cite{nouwens2020dark}. Additionally, several governing authorities, such as national Data Protection Authorities (DPAs) have been formed and have set up guidelines for GDPR compliance.  

Due to the ubiquity of consent dialogs, users in the EU and UK are now generally familiar with this process of consenting \textit{to something}, but do they actually know \textit{what} they are consenting to? The UK and EU GDPR mandates that consent be \textit{informed}~\cite{EUdataregulations2018}, yet many studies on privacy notices have shown that being informed is often simply \textit{assumed}~\cite{utz2019informed, nouwens2020dark}. 

\begin{figure}[!ht]
\includegraphics[width=6cm]{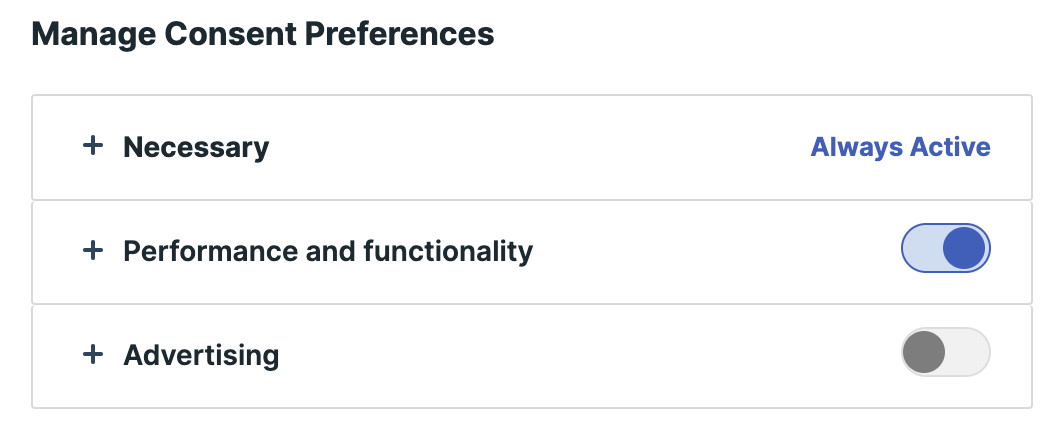}
\centering
\caption{An example of a privacy notice with its listed data collection purposes.}
\label{fig:dc_purposes}
\Description{An example of a privacy notice with its listed data collection purposes: Necessary, Performance and Functionality, and Advertising.}
\end{figure}

At the core of informed consent lie the data collection purposes for which users are sharing their data for. Examples of such purposes presented in privacy notices may include: ``Strictly Necessary'', ``Advertising'', ``Analytics'', etc. as illustrated in Figure~\ref{fig:dc_purposes}. 
Current regulatory guidelines for how data processing 
purposes should be described or named vary significantly~\cite{machuletz2019multiple, santos2021cookie}, therefore diversity exists between various websites, DPA guidelines, and CMP templates. 
Problems also exist within these purpose names and descriptions, such as: how these names do not always accurately map onto the technical services provided~\cite{bollinger2022automating}, or how, as we have observed in this paper, they can exploit cognitive biases. Some purpose names and descriptions are trickier for users to understand than others. Overall, whether purpose formulations are effective at informing users about what their data would be used for remains an understudied topic. 

%

%


Focusing on the data collection purposes and identifying the major issues with them are important for three reasons. 
First, although there is research showing that users are often the weakest point in security~\cite{sasse2001transforming, hughes2021human}, there will always be a small group of users who want to be informed about their privacy, and take their time to make privacy-conscious choices. Currently, privacy notices are riddled with deceptive designs~\cite{nouwens2020dark, machuletz2019multiple, utz2019informed, krisam2021dark}, therefore we believe that reframing these purposes will make online data collection practices more transparent and informative for these users.
Second, privacy nutrition labels have shown that when privacy information is presented to users in a digestible manner, users often are more informed without feeling overwhelmed~\cite{kelley2009nutrition, kelley2010standardizing}. Therefore, we lay the groundwork for future work on redesigning more user-centric consent systems. 
Third, reframing data collection purposes allows for more efficient management and repurposing of data~\cite{andreotta2022ai, biega2023data}, when combined with appropriate technical and legal measures. Many organizations are not fully aware of what is happening with the data they collect 
~\cite{bad_data}, which is problematic from the perspective of user privacy. 

There is a complex network of adtech vendors and third parties involved in the collection and processing of user data~\cite{veale2022impossible}, but users tend to group third parties, service providers, and other data controllers as being the same entity~\cite{kyi2023investigating}. Moreover, privacy notices are an important mechanism wherein information about purposes is disclosed to users. There is value in understanding how users perceive these purposes, and understanding how we can make them more user-friendly because users are requested to make decisions regarding these purposes every time they are online. We focused on HCI-specific contributions, but suggest that a transdisciplinary solution encompassing HCI, the law, and technical stakeholders be involved to drive meaningful change.

We thus argue that more attention needs to be paid to the data collection purposes and the text within the privacy notices to develop better formulations that accurately and effectively inform users about how their data is being processed. Hence, we investigated how users perceive the data collection purposes they might be consenting to, and identified elements of an effective purpose name and description to better improve the \textit{informed} aspect of informed consent. 

The research questions in this paper are:
\begin{enumerate}
    \item How do users evaluate common data collection purposes and their descriptions?
    \item How do users prefer data collection purposes be named and described so that they are more user-friendly?
\end{enumerate}

To answer these research questions, we carried out interviews with European internet users ($n=23$) to understand how they perceived common data collection purposes.
We found that most purpose descriptions do not contain information participants want to know, including how long their data is retained for, and how to request their data be deleted.
For purpose names, participants found some names to be more clear and understandable compared to others. Participants had varying opinions for the different purposes presented; some purposes, such as \textit{Strictly Necessary} purposes being were more accepted than \textit{Personalized Advertising} for sharing data with. \textit{Statistics and Analytics} or \textit{Performance and Functionality} purposes were not commonly understood properly. Based on our findings, combined with research insights from psychology, we propose guidelines to improve data collection purpose names and descriptions, which may likely improve how \textit{informed} users are in the informed consent process.

\section{Background} 
\label{sec:background}
We describe the legal and design foundations for our work. There are few legal specifications about what purposes data can be collected for, how they should be described, and how to best inform users before collecting their consent, which impacts the usability of privacy notices on a deeper level beyond the UI. 

\subsection{Legal Background}
\label{sec:legalbackground}
The GDPR applies to the processing of personal data~\cite{EDPB-4-07} of EU and UK users and requires organizations to choose a legal basis to lawfully process personal data (Article 6(1)(a)). When the legal basis being applied is consent, the GDPR also defines the requirements for a valid consent. Article 5(1)(a) and Recital 60 of the GDPR also require disclosure of information which is triggered by the principles of lawfulness, fairness, and transparency.
The ePrivacy Directive (ePD) provides supplementary rules to the GDPR in particular for the use of tracking technologies, such as cookies. Article 5(3) of the ePrivacy Directive requires websites to give clear and comprehensive information when requesting consent for non-necessary tracking purposes for the service requested by the use (such as targeting advertising, social networks, third-party analytics).

Some purposes are exempt from consent, such as \textit{functional} or \textit{essential} trackers (Recital 66 ePD). 
The only way to assess with certainty whether consent is required is to analyze the purpose of each tracker on a given website~\cite{EDPB-2-13, EDPB-3-13, fou-IWPE-2020}. Cookie purposes allowing website owners to retain the preferences expressed by users, regarding a service, should be
deemed essential or technically necessary. However, research has also found that sometimes purposes deemed essential are being used for non-essential purposes, such as advertising~\cite{bollinger2022automating}.

\subsubsection{Scope of the GDPR in the UK} As our participant pool consists of UK residents in addition to EU residents, we address the scope of the GDPR in the UK. As a result of Brexit, the EU GDPR is not in effect within the UK, however, the provisions of the EU GDPR were incorporated directly into the UK law as the ``UK GDPR''\cite{UK_dataprotection}. In practice, there is little change to the core data protection principles, rights and obligations, and organizations can operate as they did pre-Brexit~\cite{ico_eu}. Additionally, the UK DPA maintains that they work closely with the EU for data protection~\cite{ico_eu}.
Thus, we use the term ``GDPR'' throughout the paper to refer to both the UK and the EU GDPR.

\subsubsection{Legal requirements for purpose formulation}
Pursuant to the principle of purpose limitation (Article 5(1)(b) GDPR~\cite{EDPB-3-13}), personal data 
can be collected for specified, explicit and legitimate purposes only. 
Santos et al. studied various legal documents and systematized the legal requirements for purposes, which require 
i) \textit{explicitness} (availability, unambiguity, shared common understanding), ii) \textit{specificity}, iii) \textit{intelligible} (non technical terms, conciseness), iv) \textit{clear and plain language} (straight forward statements, concreteness), and concrete requirements for consent: v) \textit{freely given} and vi) \textit{informed} consent~\cite{santos2021cookie}. 

Most important to our work is the \textit{informed consent} requirement. Whenever tracking technologies are deployed on a user’s device, the user must be given clear and comprehensive information, and the content information must comprise the purposes of processing and the means for expressing their consent, pursuant to Article 5(3) of the ePD. 
The need to present information on the processing operations is triggered by the principles of lawfulness, fairness, and transparency depicted in Article 5(1)(a) and the recitals of the GDPR. In particular, Recital 60 explains that ``a data controller should provide a data subject with all information necessary to ensure fair and transparent processing, taking into account the specific circumstances (…).''

\subsubsection{DPAs require clear purposes}
Regulatory guidelines provide examples of purposes, yet there is no consensus on which formulation of purposes is preferred. As such, several national DPAs have different standards and guidelines for organizations subject to the GDPR. The Italian DPA confirms the absence of a standardized naming convention for cookies’ purposes~\cite{garante}.

The UK DPA acknowledges that while providing information about cookies’ purposes equates to transparency requirements, users may not always understand that information.
The UK DPA encourages websites to make an  effort to explain their activities in an understandable manner, but it does not impose strict requirements~\cite{ico_cookies}. The Latvian DPA requires that the information provided not contain unduly legal or technical language~\cite{latvia_cookies}.

The French DPA recommends formulating purposes in a descriptive and intuitive name so that users can be fully aware of the possibility of exercising a choice by purpose. It says that purposes should be formulated
``in an intelligible way, in a suitable language and clear enough to allow users to understand precisely what they are consenting to.'' It also recommends that each purpose be highlighted in a short and highlighted title, accompanied by a brief description~\cite{cnil_cookies}. 

In addition, the EDPB Taskforce~\cite{taskforce} acknowledges that some service providers classify ``essential'' or ``strictly necessary'' cookies and processing operations which would not be considered as ``strictly necessary'' within the meaning of Article 5(3) ePD or the ordinary meaning of ``strictly necessary'' or ``essential'' under the GDPR. 

\subsection{Related Work on Data Collection Purposes}


In addition to guidelines from the various DPAs regarding which purposes to use, data controllers need to consider whether to leave purposes general, therefore giving users fewer choices for control, or more specific, therefore giving users more choice. 

Utz et al.'s analysis of privacy notice interfaces found that 45.5\% of banners used generic purposes, such as ``improving user experience'', 38.6\% used specific purposes, such as ``ad delivery'', and 16.9\% did not even mention their purposes~\cite{utz2019informed}. Korff et al. found that when participants were presented with more privacy setting choices, they were less happy and more likely to regret their choice~\cite{korff2014too}. 

When presented with more specific purpose choices, Habib et al. found that users are more likely to accept only \textit{Strictly Necessary} cookies or make more granular consent choices if the UI made it easy to do so~\cite{habib2022okay}. They studied user comprehension of four purposes categories developed by The UK International Chamber of Commerce: i) \textit{Strictly Necessary}, ii) \textit{Performance}, iii) \textit{Functionality}, and iv) \textit{Targeting/Advertising}, and found that the purpose categories of \textit{Performance} and \textit{Functionality} were the most misunderstood by users~\cite{habib2022okay}. 

Previous work has shown that the design of a privacy notice impacts how users make consent choices~\cite{nouwens2020dark, machuletz2019multiple, habib2022okay}. Service providers, consent management platforms (CMPs), and third party vendors, commonly use deceptive practices to collect users' consent, such as by making it difficult to reject consent, and not properly informing users~\cite{nouwens2020dark, machuletz2019multiple, utz2019informed, krisam2021dark}. As such, many users find privacy notices to be annoying, and do not pay much attention to them~\cite{kulyk2018website}. 

Bouma-Sims et al. found that few users actually read purpose definitions, though no significant difference in comprehension was noted when definitions were provided~\cite{bouma2023us}. Kyi et al. found that users tended to be most accepting of sharing data for \textit{Strictly Necessary, Security and Debugging,} and \textit{Fraud and Law Enforcement} legitimate interest purposes, but least accepting of sharing data for \textit{Personalized Ads} and \textit{Sharing Data with Third Parties} legitimate interest purposes~\cite{kyi2023investigating}. 

Not only must organizations consider the number of purposes presented in privacy notices, but they should also consider users' perceptions of these data collection purposes. The data collection purposes offered, and their applications are often a multi-stakeholder situation, involving many different actors, such as IAB Europe and CMPs in addition to the service provider~\cite{kyi2023investigating, hils2020measuring}.

\subsection{Deceptive Design Beyond User Interfaces}
While there has been a plethora of research looking at deceptive design choices in the user interfaces of privacy notices~\cite{utz2019informed, kyi2023investigating, machuletz2019multiple, nouwens2020dark}, less attention has been paid on deceptive practices outside of the UI~\cite{kyi2023investigating, santos2021cookie}. Of the work that has looked into non-UI deceptive designs, studies have shown that deceptive practices go well beyond the UI, such as deceptive linguistic practices~\cite{kyi2023investigating, santos2021cookie, ma2022prospective}. 

Previous work by Santos et al. has found that 89\% of privacy notices violate the GDPR; 61\% were too vague in describing  purposes (thereby violating the \textit{purpose specificity} principle), and 30\% framed their data practices in positive language (violating the \textit{freely given and informed} requirements for consent)~\cite{santos2021cookie}. Kyi et al. found that linguistic deceptive designs were exploited in the use of legitimate interests, such as by providing placebic and/or positive explanations to users about their data collection practices, and being vague about legal terms~\cite{kyi2023investigating}. 

%
%
Habib et al. and Ma et al. found that loss aversion text, which is where privacy notice text might point out the negative outcomes of not accepting all cookies, was influential in making users believe they had to accept all cookies~\cite{habib2022okay, ma2022prospective}. This practice should therefore be avoided as a practice by data controllers~\cite{habib2022okay}. Berens et al. also confirmed this finding, showing in their study that the phrasing for accepting or rejecting cookies can influence users' behaviours~\cite{berens2022cookie}. 

Related work in this space suggests that descriptions and linguistic elements, such as positive or negative framing and being transparent about practices, can impact users' consent choices and perceptions. We extend upon this work by looking at the linguistic elements (i.e., purpose names and descriptions) of privacy notices, and the challenges and opportunities for consent that they may present.

\section{Methods}
We conducted semi-structured interviews with 23 English-speaking participants over 18 years old from the UK and Ireland. We recruited participants using Prolific\footnote{\url{prolific.co}}, a website for recruiting online participants for research studies. Since many languages are spoken within the countries scoped by the GDPR, and different languages might present different nuances in the text of a privacy notice, we recruited from the UK and Ireland to increase the chances that participants were exposed to English-language privacy notices, and thus more familiar with the data collection purposes we presented to them. 

Interviews took approximately one hour to complete, after which participants were compensated \texteuro{23} for their time. Our Institutional Review Board declared that this study was exempt from ethical review, however we had to collect consent from participants for the audio recordings to be GDPR-compliant. Interviews were conducted during June 2023. After the interviews were conducted, we used an automatic tool to transcribe our interview audio, and two researchers annotated the interviews. As our study spans across multiple disciplines, we had a multidisciplinary team, consisting of those from computer science, psychology, HCI, and legal backgrounds.


\subsection{Selecting Data Collection Purposes}
As purposes are flexible, and there is no standardized list of  purposes and their descriptions, we had to search through various DPA guidelines, Consent Management Platforms (CMPs), and the ePrivacy Directive to collect purposes and descriptions for the second and third sections of our interview. 
Regulators propose various ways to formulate purposes. Some DPAs, such as those discussed in Section~\ref{sec:legalbackground}, provide more concrete guidelines about how to name and describe purposes, which has helped to guide our study.

We initially collected a set of 201 purposes and descriptions from 6 CMPs (OneTrust, Quantcast, TrustArc, Cookiebot, LiveRamp, and Crownpeak) and 39 companies from our own online browsing and looking at previous literature studying data collection purposes~\cite{hils2020measuring, habib2022okay, santos2021cookie}. We ultimately decided on a smaller set of six purposes for conducting the interviews because we noticed that most purposes we collected were CMP- and DPA-based purposes. Therefore, focusing on CMP- and DPA-based purposes provided a wider coverage of purposes and descriptions that are used in practice~\cite{hils2020measuring}.

The six data collection purposes we decided on are: 
\begin{enumerate}
    \item  \textit{Strictly Necessary / Essential / Required};
    \item \textit{Performance / Functionality};
    \item \textit{Statistics / Analytics};
    \item  \textit{Advertising};
    \item  \textit{Personalized Advertising}; and
    \item \textit{Personalized Content}
\end{enumerate}
The different descriptions corresponding to each purpose are included in Section 2 of our Supplementary Materials. 

We decided on these six purposes because we wanted purposes that are both widely used and broad enough that they covered a variety of different uses. As a point of comparison, some of these purposes have been studied in other papers~\cite{habib2022okay, kyi2023investigating}, but we also added other purposes that have not been studied yet to gain new insights. While \textit{Advertising} and \textit{Personalized Advertising} seem similar, we wanted to see whether participants could differentiate between the two, and how they felt about many purposes with different names being related to advertising. 

We pilot tested our interviews with three participants; all regularly used the internet in English, one participant was from a computational background, one from a non-technical privacy background, and one from a non-technical background for variety. The pilot tests helped us reformulate our interview questions, and indicated that showing participants six data collection purposes, each having between three to five different descriptions, was within participants' attention limits. 

\subsection{Interview Procedure}
During the interviews, we first introduced our work and the research team, then asked participants to fill out a consent form and a demographics form. Thereafter, we gave participants definitions of what a ``data collection purpose'' meant, and provided screenshots of data collection purposes in privacy notices to provide more context. We then proceeded with the interview questions, which we summarize below (see Section 1 of our Supplementary Materials for the full interview protocol). All interviews were conducted by the first author on Zoom, with the option for participants to turn on their video.  

Our semi-structured interview consisted of three sections. 
In the first section, participants were asked about what they expect organizations to disclose in privacy notices (Q1.1), whether they think it is necessary to share their data with organizations (Q1.2), how they feel about privacy notices telling them of services they will miss out on if they declined cookies (Q1.3), and how well-informed they feel about online data practices (Q1.4). 

In the second section, participants were presented with the names of six data collection purposes without any definitions (\textit{Strictly Necessary / Essential / Required, Performance and Functionality, Statistics and Analytics, Advertising, Personalized Advertising,} and \textit{Personalized Content}). For each purpose, we asked what they think happens with their data under this purpose (Q2.1), whether they would feel comfortable sharing data for that purpose (Q2.2), and what they think would happen if they denied consent for that purpose (Q2.3). We repeated this set of questions for each purpose.

In the third section, participants were given a link to a document that showed each of the six data collection purposes presented in section two of the interview, and presented with various descriptions from different sources to read (see Section 2 of our Supplementary Materials for these descriptions). After participants read the description for one purpose, they were asked how they would describe that purpose in their own words (Q3.1), how similar they felt the descriptions were (Q3.2), whether there was a description they preferred the most (Q3.3), how well-informed they felt about how that purpose uses their data (Q3.4), what could be improved in the descriptions (Q3.5), how clear the purpose name was in describing what it does (Q3.6), and if the name was not clear, whether they had suggestions for a better or improved name (Q3.7). We repeated the procedure of having participants read the descriptions and answer these questions for each of our six purposes. 

\subsubsection{Participants}
We recruited participants who were over 18, spoke and used the internet primarily in English, and lived in the UK or Ireland to ensure exposure to English privacy notices. Despite Brexit, the UK GDPR remains largely similar to the EU GDPR, and UK residents are still subjected to responding to privacy notices~\cite{ico_consent}. Therefore, UK residents should be just as familiar with privacy notices as EU residents, and including UK residents would give us access to a larger pool of potential research participants. We stopped recruiting at 23 participants because we reached saturation by this point, meaning we stopped hearing new topics being brought up at this point~\cite{glaser2017discovery, saunders2018saturation}. Most interviews reach saturation between 9 and 17 interviews, and for studies with a general population, such as ours, saturation is reached at approximately 16 interviews~\cite{hennink2022sample}.

Our participants were roughly split between men (48\%) and women (52\%), all primarily used the internet in English, and all participants except for one had been living in the UK or Ireland for over four years. We did not collect information about participants' educational backgrounds, but did aim for a representative sample across gender and age, therefore we had a wide variety of ages represented in our sample; 17\% were between the ages of 18 to 24, another 17\% were between 25 to 34 years old, 26\% between 35 to 44 years old, 23\% between 45 to 54 years old, and 17\% were over 55 years old. Most of our participants came from non-computational backgrounds. 

\subsection{Data Analysis}
\textbf{Codebook.} Our qualitative interview data was analyzed by two authors of this paper, starting with an inductive, open-ended approach to data analysis, then a deductive approach with a codebook that was revised during the annotation process~\cite{swain2018hybrid}. This method was preferred since the interview investigated different data collection purposes, allowing for us to connect codes to specific purposes we studied. See Section 3 of our Supplementary Materials for our codebook.

\textbf{Annotations.} Since the first annotator conducted all the interviews, they initially open-coded three interviews to form an initial codebook. Afterwards, the first and second annotator met to annotate another three interviews together, discuss, and adjust the codebook as needed. The annotators then coded the same set of another six interviews separately, meeting after each interview to compare codes and discuss. After they reached an interrater reliability (IRR) of 80\% ($kappa$ = 0.79), which indicates substantial agreement~\cite{mchugh2012interrater}, the annotators split the rest of the interviews and annotated them separately. For increased validity, the annotators individually re-annotated the initial interviews where IRR was not yet reached. 


\section{Results}
In this section we describe our qualitative results. As our interviews were not directly measuring quantities, we have used terminology present in previous qualitative HCI studies to give estimates of quantities~\cite{habib2022okay, emami2019exploring, kyi2023investigating}.

\begin{figure}[!ht]
\includegraphics[width=7cm]{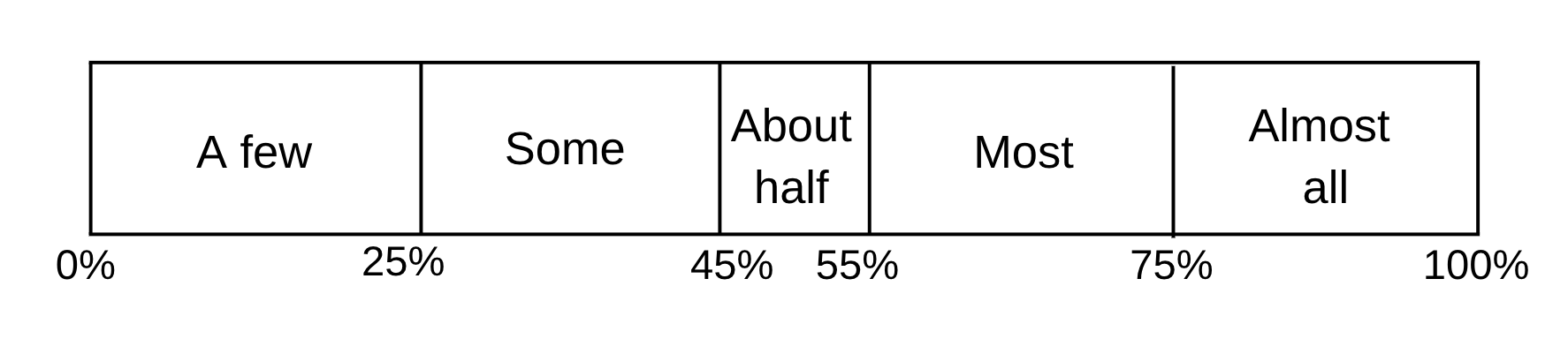}
\centering
\caption{Terminology used to represent the frequency of themes in our qualitative results. This graphic was taken from ~\cite{kyi2023investigating}.}
\label{fig:frequencieswords}
\Description{A figure representing terminology used to describe the frequency of themes in qualitative research. A few refers to approximately 0 to 25\%, Some refers to approximately 25 to 45\%, About half refers to approximately 45 to 55\%, Most refers to approximately 55 to 75\%, and Almost all refers to 75 to 100\%.}
\end{figure}

\subsection{Perceptions of Data Collection Practices}
\label{sec:perceptions}
During the interview, participants provided comments about their thoughts regarding data collection practices in general. 

\textbf{Participants did not feel informed about data collection practices.}  
%
None of the participants felt well-informed of online data practices, even if the law prescribes the variety of information an organization  needs has to disclose to users to ensure fair and transparent processing (Articles  5(1)(a), 14, Recital 39 GDPR).

Almost all participants said they did not know i) what data is being collected, ii) how it is collected, iii) why their data is collected, iv) to whom this data is being sent to, nor the v) sensitivity of the data being collected. Additionally, many participants said they do not trust the information websites share about how they process user data, believing that all purposes were being used covertly for advertising in some way. Accordingly, mandated informational and transparency requirements are not efficient for meaningful choices. 

%
%

\textbf{Participants saw privacy notices negatively.} 
Very few participants read privacy notices regularly; almost all participants felt privacy notices were annoying and just usually do what it takes to get rid of them quickly, such as clicking ``Accept all'' or accepting only necessary cookies by default. This echoes findings from previous research which found that users tend to not interact with privacy notices very thoroughly~\cite{nouwens2020dark, habib2022okay, bouma2023us}.


\textbf{Participants believed that sharing their data is unnecessary, or a trade-off.}
When asked how necessary participants thought it was to share their data with organizations, we received a mixed response. Some participants believed it was necessary to share as much data as organizations are collecting in exchange for using these free services. 

However, many others believed it was not necessary to share as much data as organizations are (perceived to be) collecting because these are often over-collecting data. As one participant stated, \textit{``I don't think it's necessary, but it's just becoming more the norm nowadays. Most people just don't really know that actually their data is being used or stored at all''} (P2). 

Some participants said that while sharing data is not necessary, they feel resigned to share it because organizations have more power in the end. They felt they have little control over the data they can share with organizations. Even if users wanted to decline sharing data, most believed organizations would still find a way to collect their information, echoing a finding from Kulyk et al.~\cite{kulyk2020has}.

\subsection{Perceptions of Data Collection Purposes}
Herein we discuss participants' general perceptions of the six data collection purposes we presented: i) \textit{Strictly Necessary / Required / Essential}, ii) \textit{Performance / Functionality}, iii) \textit{Statistics / Analytics}, iv) \textit{Advertising}, v) \textit{Personalized Advertising}, and vi) \textit{Personalized Content}. 

\textbf{Most participants believed nothing would happen if they declined tracking.} When asked what they think would happen if they declined tracking for a purpose, most participants believed it would make no noticeable impact on their current online experience. This finding is somewhat contrary to previous work which showed that only a few participants believed nothing would happen by declining tracking~\cite{kulyk2020has}. This difference may be due to recent case-law compelling companies to present more balanced options in consent notices, so that users reject tracking more easily in consent notices and notice not much happens when tracking is declined~\cite{cnil_optout}.
As P23 described, \textit{``I have declined them sometimes, but I've never noticed any difference between declining them or accepting them.''} Other participants' quotes supplement this finding: \textit{``I don't think anything would happen really, except maybe you'd end up with less spam''} (P2) and \textit{``I think nothing would probably happen, except that the provider would probably be less than happy because you'd be less of a useful customer to them''} (P9).

Some others believed they might get fewer services relating to the particular purpose they declined, such as less targeted advertising if they declined \textit{Personalized Advertising}, or would not be able to access the site depending on the purpose(s) they declined. A few participants said they lacked the technical knowledge to assess how site functionalities would be impacted if they declined cookies, therefore felt the need to accept all tracking. 

\textbf{Participants either felt threatened or inquisitive about missing services because they rejected tracking.}
When asked how they felt about privacy notices that tell them they are missing out on certain services by rejecting tracking, we had mixed responses from participants. Some participants found this information important, and wanted to know about the services that they would be missing out on if they declined purposes. 

Conversely, some believed they did not need to know what they were missing out on and felt these messages were threatening, as if organizations were forcing them to share their data. As stated by P4, \textit{``It feels like a manipulation, if I'm honest. It's kind of a way of them trying to get you to consent to things.''}

\textbf{Most participants believed \textit{Strictly Necessary} purposes were mandatory, but some participants had doubts.} 
Most participants understood \textit{Strictly Necessary} purposes as being mandatory, and that they have to accept it 
to access a website. 
However, we also had some participants who doubted whether this purpose was actually necessary for site access. This is a correct assumption according to Article 5(3) ePD. The assumption of necessity requires a technical analysis to assess whether any tracker is indeed necessary for a website to work~\cite{bollinger2022automating}. 

Additionally, some participants believed this purpose was a vague, ``catch all'' kind of purpose. As explained by P7: \textit{``I don't think it's necessary, but it does seem that most of the defaults are to allow everything. On a lot of sites, it does seem where it's only strictly necessary cookies (being shown to users). So it seems like they they're just relying on people just to click `accept' without really looking at any of the other details.''}
%
Participants sometimes described that this purpose provided advertising or provided a better site experience for users, conflating it with other purposes such as \textit{Performance and Functionality} purposes.
 %
 
\textbf{Participants linked the majority of purposes to advertising.}
Most participants correctly thought \textit{Performance and Functionality} purpose provided them with more effective services and remembered their choices. 
Yet, some participants also believed this purpose provides them with targeted ads, as explained by P22, \textit{``I would have guessed this similar to \textit{Statistic and Analytics}, which I would like to call `advertising.' I don't know what I would expect to be different.''}

\textit{Statistics and Analytics} purpose were commonly believed to be used to analyze data from users using the site, and to analyze website statistics for future improvements. In line with perceptions from other purposes we presented, some participants believed \textit{Statistics and Analytics} was yet another purpose being used for sending marketing and advertising materials to users.

Some participants conflated \textit{Personalized Content} with \textit{Personalized Advertising} purposes. 
However, most participants believed \textit{Personalized Content} was meant for creating user profiles and showing personalized content based on these profiles. Regarding the creation of user profiles, many participants said they wanted more information from organizations about how they were creating their profiles. As explained by P19, \textit{``It just tells me that they will give me content based on what I like, it doesn't tell me anything about how my data is used.''}

\textbf{\textit{Advertising} was conflated with \textit{Personalized Advertising}.} In the collection of purposes presented, we found that certain DPAs (French and Spanish DPAs), and CMPs (OneTrust, TrustArc, and LiveRamp) differentiated between \textit{Advertising} and \textit{Personalized Advertising} purposes. \textit{Personalized Advertising} and \textit{Advertising} purposes were always confused with each other by participants, wherein they believed that both of these purposes were meant to serve personalized ads. In reality, \textit{Advertising} is used for delivering generic ads to users. All participants were correct in their interpretation of \textit{Personalized Advertising} purposes, believing it is used to serve users personalized ads. 

\textbf{Participants were not comfortable with sharing data for \textit{Advertising} purposes.} As for user comfort with sharing data for the six purposes we presented to participants, it differed by purpose. Most participants were not happy sharing data for \textit{Personalized Advertising} and \textit{Advertising} purposes. Instead, participants said that they were more comfortable sharing their data for \textit{Strictly Necessary}, \textit{Performance and Functionality}, and \textit{Statistics and Analytics} purposes. When they found out what \textit{Personalized Content} purposes did, and realized it was not the same as \textit{Personalized Advertising}, most participants said they were comfortable sharing their data for this purpose due to the perceived convenience of this purpose, echoing previous research findings~\cite{kyi2023investigating, kozyreva2021public}.

\subsection{Perceptions of Purpose Names}
When asked how clear the name of a given purpose was, participants had some mixed responses regarding certain purposes. 
When it came to suggesting concrete names to make them more comprehensible, participants were generally much better at knowing why a name was unclear rather than suggesting a new and improved name.

\textbf{\textit{Personalized Advertising} is a clear name.} This name was the clearest purpose name to participants compared to the other purposes we presented. All of our participants felt the name was expressive 
of what the purpose does, and participants' perceptions of the purpose matched what purpose descriptions said.
On the flipside, almost all participants conflated \textit{Advertising} for \textit{Personalized Advertising} purposes, believing them to provide the same functions.


\textbf{Some purposes make use of conjunction and synonymy.} Sometimes, purposes are called by several names by different companies, DPAs, or CMPs, such as \textit{``Strictly Necessary / Required / Essential''}. Additionally, purposes may also use conjunctions, which is when several functionalities are combined into one purpose, such as \textit{``Statistics and Analytics''} or \textit{``Performance and Functionality.''} 
Notably, regulators also use synonyms to name purposes, such as using ``Technical / Required / Functional cookies'' as per the Latvian DPA's guidelines~\cite{Latvian-guidelines-2022}. Conjunctions, such as ``Personalised ads and content, ad and content measurement, audience insights and product development'' are used by CMPs like LiveRamp. 
 
%
Where conjunction names were presented, participants often had a preference for one of these names, finding it clearer and/or more transparent than the other(s), as explained below.


\textbf{\textit{Strictly Necessary / Required / Essential}}. Some participants wondered if these purposes were actually ``strictly necessary'' for a website to work.  
In the words of P6, \textit{``I think it (the name, \textit{Strictly Necessary / Required / Essential}) needs to be improved because `Necessary for who?' is my question. It might be necessary for the person that wants your information, but not to me as a person using it. I would like a little bit of clarification on who it's necessary for because from the title it looks as though it's just necessary for the organisation and not for me.''} 


\textbf{\textit{Statistics and Analytics}}. Some participants found \textit{Analytics} to be more transparent and descriptive compared to \textit{Statistics}. As described by P18, \textit{``I would go with `Analytics' rather than `Statistics'. For me, it's just knowing how they sort of use the data. I don't need to know the figures and stuff.''}

\textbf{\textit{Performance and Functionality}}. Most participants found \textit{Performance} to be a better name. To many participants, \textit{Functionality} implies that this purpose is important, or even necessary, for the website to function, whereas \textit{Performance} aligns better with the descriptions of this purpose, which relates to UX improvements and site experience. As P20 explains, \textit{```Functional' to me just means that it makes it (the service) work, whereas `Performance' indicates that it will try to be as best as it can, like how good is the performance.''}

\subsection{Perceptions of Purpose Descriptions}
We presented multiple descriptions from various sources (the ePrivacy Directive, DPAs and CMPs) for each purpose. Based on participant responses after viewing these descriptions, we gained insight into what can be done to improve purpose descriptions.

\textbf{Participants preferred simple descriptions with less technical jargon.}
Most participants preferred simpler descriptions, but with more relevant and concise information about how their data is processed, as described in Section~\ref{sec:perceptions}. Further, they preferred when descriptions avoided technical and legal jargon, such as terms like ``persistent cookies'' and ``SSO''. As P15 says, \textit{``It talks about single sign on (SSO), and I feel like, like... as someone who doesn't really know this term, I wouldn't really want to see the `SSO' part. I wouldn't want to make it seem as though it's harder to read than it actually is.''} 
Consent choices in current interfaces often employ language that is too technical or confusing for users~\cite{Fatemeh-knowledgeOnTracking}, which contradicts legal requirements that mandate consent collection to be understood by an ``average member of the intended audience''~\cite{Transparency29WP, EDPB-3-13}.

%

\textbf{Participants suggested that privacy notices should be more visually appealing to capture attention.}
To improve readability and capture user attention towards data collection purposes and descriptions, some participants suggested that privacy notices describe purposes in point-form, and made better use of colour and icons to make them more appealing, rather than the current paragraphs of text describing purposes. 

%

\textbf{Participants want to be further informed about data retention and data rights.}
In addition to being better informed of what data is being collected for and who it is sent to, participants also wanted to be informed of the time their data is retained for once they accept tracking for a purpose, what happens to their data once they end the session, and the sensitivity of the data that is being collected about them. 

Participants also mentioned wanting to know more about 
how they can request organizations to have their data deleted, what happens to their data if they reject cookies, and the services that are still provided once they reject cookies. As described by P3, \textit{``(I want to know) how I can go and remove it (my data) if I've already consented to it. It would be nice if these parts of data collection were disclosed here (the privacy notice) as well.''}



\textbf{Participants wanted more reassurance in the description of purposes.}
Most participants preferred when purpose descriptions provided them with reassurance that their data would be kept safe, such as when descriptions mention that user data would be kept anonymous, or not be used for profiling purposes. Given the lack of transparency in how user data is handled, participants preferred it when companies gave them reassurance about keeping their data private. As P15 puts it, \textit{``It mentioned they don't directly store personal information, which is quite nice to know.''}

The need for reassurance was also observed by some participants who said they would like organizations to better describe the technical elements behind their data collection, such as by providing hyperlinks to find more information, or providing brief definitions for technical terms. Despite these suggestions, participants also said they would be unlikely to read or find out more about technical terms, but they want to feel reassured to know that there is the option to find out more information if they chose to.
The EDPB guidelines~\cite{EDPB2023} match user perceptions suggesting the use of technical definitions and examples. 
%

\textbf{Descriptions for the same purpose can be perceived as dissimilar.}
When presented with several different descriptions for the same purpose, participants often felt descriptions were different due to three reasons. 
First, participants indicated a disparity in the depth of information given; some descriptions were more informative than others, therefore were perceived to be more transparent. 
Second, a difference resided in the examples purpose descriptions gave. Participants noted that the examples of services provided would vary, making them confused about what the purpose actually did. 
Last, participants sometimes commented that these different descriptions were sometimes giving them conflicting information. 
This was especially prominent in \textit{Statistics and Analytics}, and also \textit{Performance and Functionality} purposes, a finding Habib et al. also found~\cite{habib2022okay}. 
As demonstrated by P3, ``I feel like they have slightly different definitions (for \textit{Performance and Functionality)}. It kind of takes me back to \textit{Statistics and Analytics} a little bit where we were talking about the performance over there as well.'' 


\textbf{Different purpose descriptions can be perceived as describing the same functions.}
Participants sometimes mentioned that, based on descriptions for different purposes, some purposes seemed similar because of the services mentioned and perceived functions of that purpose. 
For example, some participants pointed out that two descriptions for \textit{Statistics and Analytics} mentioned advertising, even though it is not an advertising purpose. Similarly, participants commonly confused \textit{Statistics and Analytics}, \textit{Performance and Functionality}, and \textit{Strictly Necessary} purposes with each other, believing their descriptions sometimes overlapped. 

\section{Discussion}


In this paper we investigated two research questions: i) how users evaluate commonly used data collection purposes and their descriptions, and ii) how users prefer data collection purposes be named and described so that they are more effective. 

We expand on the knowledge of informed consent by focusing on ways that users can be better informed about online data collection purposes and their descriptions. 
Based on our findings, in addition to insights from research in psychology, we present several recommendations to further improve the framing of data collection purposes towards informed consent.

\subsection{Towards User-Informed Purpose Names}
Based on previous research and our findings, we make the following recommendations for reframing data collection purpose names. 

\begin{figure}[!ht]
\includegraphics[width=7cm]{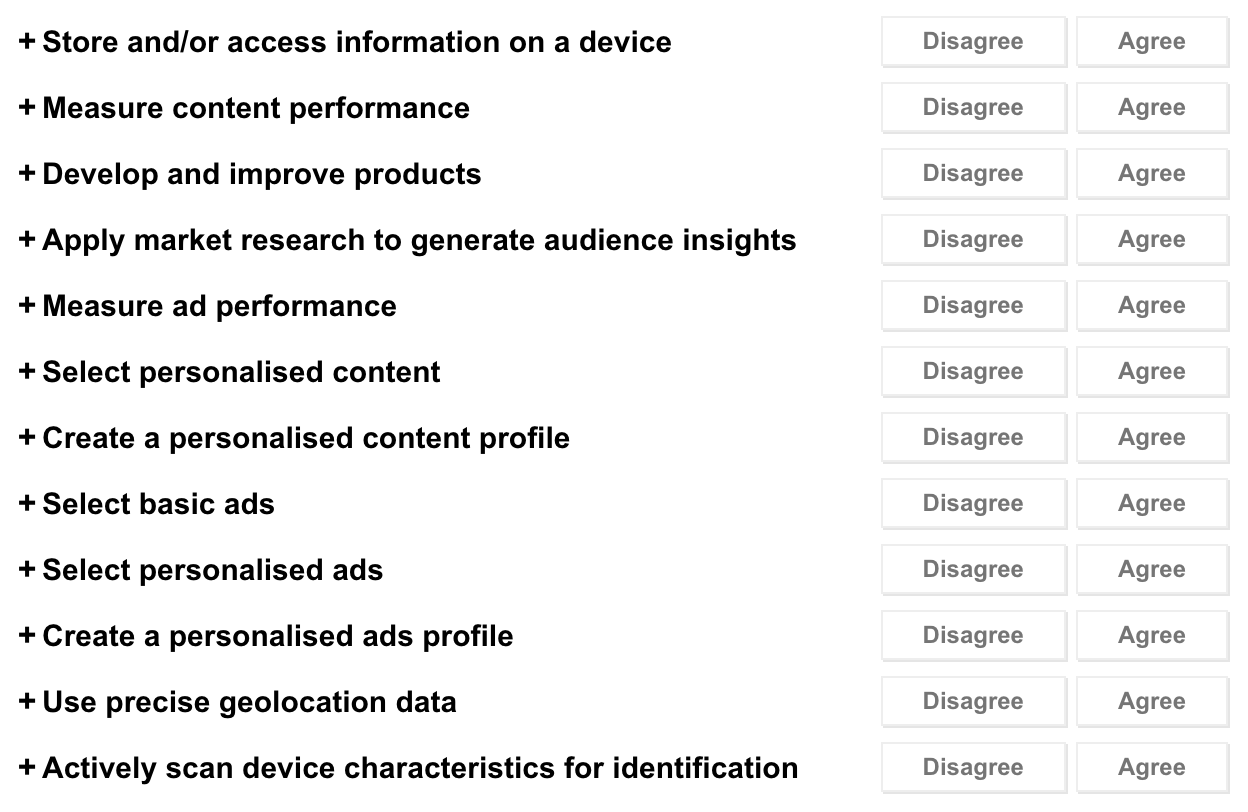}
\centering
\caption{An example of a privacy notice with very specific purposes, taken from the wild.}
\label{fig:specific}
\Description{An example of a privacy notice with very specific purposes. This example displays 12 data collection purposes which users must opt out of individually.}
\end{figure}

\textbf{A middle ground is needed between overly broad and overly narrow purposes.}
We caution against using overly broad and overly narrow purposes, which Machuletz et al. also suggested in their work~\cite{machuletz2019multiple}. Websites and regulators need to find a middle ground among the various purpose names. 
Overly broad purposes, such as presenting only \textit{Strictly Necessary} and \textit{Non-essential} purposes were deemed too vague and broad for users to fully understand what is happening with their data, and gives users less choice about their data. This is used in the case of the Italian DPA that only distinguishes between two broad categories of ``technical cookies'', and ``profiling cookies''~\cite{garante}.
Overly narrow purposes, such as when users are presented with very granular and specific purposes, overwhelm users by giving them too many options (see Figure~\ref{fig:specific} for an example). This is the case of the French DPA suggests several granular purposes in its guidelines~\cite{cnil_cookies}.
Singh et al. have shown that users prefer having three purpose options consisting of \textit{Required}, \textit{Functional}, and \textit{Advertising} cookies~\cite{singh2022cookie}. 

\textbf{Purpose names should be more specific about who they benefit.} In the case of \textit{Strictly Necessary / Essential / Required} purposes, some participants were skeptical about whether they were ``necessary'' for the user, or for the organization. As such, we recommend that organizations be clearer and more transparent about who these purposes benefit. For instance, this purpose could be renamed to \textit{``Essential for basic website functions''}, or \textit{``Required by the company''} to avoid potentially misleading users.  

\textbf{\textit{Advertising} and \textit{Personalized Advertising} purposes need to be better differentiated.} Almost all participants believed \textit{Advertising} purposes, which are for non-targeted advertising, were the same as \textit{Personalized Advertising} purposes. Therefore, we recommend that \textit{Advertising} purposes clarify that this is for non-targeted advertising purposes to avoid confusing users. For example, the name could be changed to \textit{``Non-Personalized Advertising''} to indicate the difference.

\textbf{Purpose names should avoid conjunctions.}
Research in psychology has proven that when various items are grouped together, forming a conjunction, users will often remember the first and/or last items best, a phenomenon known as the \textit{serial position effect}~\cite{murdock1962serial}. To mitigate the effects of this phenomenon, we recommend that data controllers only present one purpose to users at a time instead of grouping them together. As posited by Santos et al.~\cite{santos2021cookie}, the use of bundling infringes upon the purpose specification principle. 

\subsubsection{Some purpose names were preferred over others} In our study, we found that participants often preferred one name over another in the case of purposes which were grouped together, as explained below. Therefore, we recommend that organizations consider using the preferred names instead of conjunctions for these purposes. 

\textbf{\textit{Statistics and Analytics}.}
Participants preferred \textit{Analytics} because it is more straightforward about what is happening to their data, whereas \textit{Statistics} is more vague and sounds more technical. Participants felt saying ``statistics'' or ``analytics'' on its own without providing context about what organizations were collecting data for was misleading. 

\textbf{\textit{Performance and Functionality}.} Participants preferred \textit{Performance} because it fit more with their perceptions of this purpose, believing it is meant to improve the site experience. \textit{Functionality}, on the other hand, sounds like it is a purpose necessary for the website to function, which was deemed to be misleading. 

\textbf{\textit{Personalized Content}} The name was not intuitive for some participants who believed it meant the same as \textit{Personalized Advertising} because ``content'' could also refer to advertising content. This misunderstanding could stem from the fact that most users are attuned to how organizations use data for delivering targeted ads~\cite{kyi2023investigating, kozyreva2021public, habib2022okay}. Therefore, it is recommended that the name be modified to help users better differentiate it from \textit{Personalized Advertising}, such as \textit{Personalized [insert application, e.g., Video/Search] Recommendations}. 

%

\subsection{Towards User-Informed Purpose Descriptions}


\textbf{Descriptions are lacking crucial information users wish to know.}
Our findings suggest that purpose descriptions need to be more transparent about what organizations are doing with users' data. Participants indicated wanting to know more about i) how long their data will be retained for, ii) how to go about deleting their data and reversing their previous consent decisions, and iii) the sensitivity of the information organizations are collecting from users. 



When descriptions mentioned they were personalizing ads and content or creating user profiles, many participants wanted more transparency from organizations about how their profiles were built and used. This is especially important as user data is often used to train AI models~\cite{andreotta2022ai}, and some organizations, such as Zoom, are updating their Terms of Service and not allowing users to refuse data collection for training AI models~\cite{zoom, zoom_ai}. 

Users want to know how their data is being used by organizations, and therefore more transparency about how personalization works needs to be conveyed to users. A UK Data Protection Authority (DPA) report found that when participants were informed about how the adtech system worked, participants were less likely to accept seeing online advertisements, revealing how being informed can change users' attitudes towards data sharing practices~\cite{ico_adtech}. 

We denote that the law mandates information disclosures about the risks and consequences of processing purposes. Under the legal requirement of \emph{informed} consent and the transparency principle, personal data processing must be handled in a transparent manner in relation to the user (Article 5(1)(a)), including obligations for websites to inform users about the types of data processes, data recipients (Article 14), the scope, consequences,~\cite{Transparency29WP} and \emph{risks} in relation to the processing of personal data (Recital 39). Offering users legal information for consent to tracking empowers them, but also may induce negative impacts, as receiving such information may decrease one's perception of risk~\cite{STRYCHARZ2021106750}.

Our study upholds these legal requirements as we found that users also want to have such information. Both the law and user's intentions are aligned in the sense that both assume an informed and rational user that acts deliberately towards privacy-friendly options. In practice, users may not read such information and act against their own intentions, presenting a \textit{privacy paradox}, where users' intentions do not match their behaviours~\cite{barth2017privacy}. 

\textbf{Descriptions should avoid using loss aversion language.}
Participants preferred when purpose descriptions gave examples of services the purpose provides, but sometimes found it threatening when descriptions used loss aversion language by saying they would miss out on certain services by denying cookies. When information is framed negatively, it may put pressure on users by exploiting loss aversion~\cite{Acquisti2017nudges} and nudge them towards consenting~\cite{santos2021cookie}, especially when it is unclear which functionalities will be lost.

Negative framing may nudge users towards the website's wishes when the information about what is missing is omitted, hence violating a freely given consent requirement. Bongard et al. showed that users may develop incorrect mental models of the consequences of (not) consenting to data collection and processing~\cite{Bongard2021manipulated}. We suggest that instead of using loss aversion language, descriptions could instead mention the services that are still provided if users reject their consent.

\textbf{Participants want reassurance from organizations.}
Another commonly occurring theme within our findings are that participants want \textit{reassurance} from organizations in their purpose descriptions. For example, participants said this can include providing them with more information to find out more about a technical aspect of data collection if they wanted (through a link or expandable definition), or more commonly, they wanted reassurance from organizations that their data would be kept private and secure, such as how it be anonymized, would not be used for profiling, and not shared with third parties, which is a finding in line with previous research~\cite{kulyk2023people}. 

However, we caution the use of reassurance in purpose descriptions; in some cases, reassurance is only a half-truth, such as in the case of \textit{Marketing and Advertising} where one description said ``We and third party companies / our partners use trackers for the purpose of measuring the audience of advertising on the site or application, without profiling you.'' In this case, it is true that the service provider (website) is not profiling users, but the third parties that data is being sent to are indeed profiling users, making these descriptions not fully accurate to users~\cite{termsfeed, webdev}. Therefore, descriptions need to be careful in presenting accurate information when they wish to reassure users. 


\subsection{Improving Purposes: a Step Towards Improving the Consent Ecosystem}
This study focused on users' perceptions of data collection purposes and descriptions that are commonly used in consent notices. 
We suggest that more focus be paid on combining our recommendations along with other broader consent ecosystem changes to further improve informed consent. Deceptive practices, such as only changing the purpose names and descriptions without changing the underlying consent ecosystem or data collection procedures, can undermine efforts towards truly informed consent. Ideally, changes to the online consent ecosystem would combine user-informed recommendations, along with appropriate technical and legal measures to prevent deceptive practices from taking advantage of users.

We foresee that informed purposes can be achieved through two methods: by implementing a consent nutrition label, and by adopting informed consent practices from pre-existing consent applications to further improve the \textit{informed} aspects of informed consent. 



\subsubsection{Consent Nutrition Label}
To account for the lack of user attention on security and privacy, researchers have tried implementing other ways of conveying this information~\cite{kelley2009nutrition, kelley2010standardizing, emami2020ask}. Privacy policies, which are often considered to be difficult and boring to read~\cite{schaub2015design}, have been improved through the use of privacy nutrition labels which use a tabular format to enhance user comprehension of an organization's privacy policy. They have been shown to increase information finding, and allow for users to make better comparisons between policies~\cite{kelley2009nutrition}. By leveraging better UI design, icons, and colours, users are more likely to pay attention to privacy policies, preferring them over traditional privacy policies, thus making privacy policies more accessible to users~\cite{kelley2009nutrition, kelley2010standardizing}. 

A finding from our study was that, similar to privacy policies, few participants read consent notices, finding them to be boring, difficult to read, and annoying, confirming previous studies~\cite{utz2019informed, kulyk2018website}. Therefore, a possible use case for privacy nutrition labels is to apply them to the description of purposes within consent notices to make them more accessible and enjoyable to interact with. 
The suggestion of leveraging design elements in consent notices is aligned with the GDPR (Article 12(7)) which prescribes that information disclosed to data subjects may be provided in combination with standardised icons in order to give a meaningful overview of the intended processing in an easily visible, intelligible, and clearly legible manner.

\subsubsection{Learning from Other Consent Applications}
The current state of data collection purposes and their descriptions contribute to making online consent largely uninformed~\cite{machuletz2019multiple, utz2019informed, nouwens2020dark, flanaganredesigning}. We posit that much can be learned from other, better-established areas where consent and data collection information is conveyed, such as in human subject research and healthcare settings. In these fields, informed consent is conveyed to users in a variety of formats which might translate effectively to consent notices, as supported by Andreotta et al.~\cite{andreotta2022ai}.

\textbf{Human subject research.}
In human subject research, there is often an ethics review board (ERB) that requires researchers specify the study's data handling procedures in the ethics form, in addition to a consent form that participants are required to read and sign. Data handling procedures are specific and detailed, and often listed in the consent form and/or orally conveyed to participants. 

Our study found that purpose descriptions are lacking in crucial information that most participants wanted to know more about. As such, purpose descriptions could learn from the ethics review process for human subject research and include information that participants want to know, such as how long their data will be retained for, how users can request their data be deleted, and who will have access to user data.

\textbf{Healthcare.}
In healthcare settings, patients are informed about the risks, benefits, and alternatives for a medical procedure~\cite{shah_consent, andreotta2022ai}. 
Similarly, in online consent, users should be made aware of the risks, benefits, and alternatives for consent to an organization's consent terms. Currently, users are often only presented with the benefits when they consent~\cite{santos2021cookie, kyi2023investigating} or the negatives of rejecting consent~\cite{habib2022okay, ma2022prospective, berens2022cookie}. 

\subsection{Limitations}
We only conducted our interviews with fluent English speakers who were living in the UK and Ireland to ensure exposure to English-language privacy notices. Therefore, there may be linguistic nuances and other differences that might be present in privacy notices written in other languages. As we only focused on commonly-used data collection purposes used in the EU/UK, it is also possible that there are other purposes and language nuances presented in other jurisdictions we did not capture 
in this study. Hence, we do not generalize our findings to non-English languages or other data privacy laws due to these potential differences.

On a related note, privacy notices and the law are constantly changing~\cite{Degeling_2019}. As such, we may not be able to capture all the recent changes in the data collection purposes or descriptions we showed participants, nor all the purpose name and description variations used in privacy notices. 

Our paper does not fully address the complex adtech and multi-actor nature underlying privacy notices. We conducted our study on users’ perceptions of what is presented in consent notices, because this is what users are seeing first-hand. 
Consent notices are one of the only ways in which information about how user data may be processed is disclosed to users. Our findings contribute to the field of HCI, but a transdisciplinary solution that involves relevant stakeholders in the consent ecosystem is necessary to enact change.

Lastly, participants' perceptions of, and suggestions for improving purposes and their descriptions may not line up with how they would actually act in real life, a phenomenon called the \textit{privacy paradox}~\cite{barth2017privacy}. This is why designing privacy notices void of deceptive designs and using user-friendly language is important to prevent users from choosing the easiest, deceptive options. We conducted a qualitative study to understand user perceptions of data collection purposes, laying the groundwork for future quantitative research.

\subsection{Future Directions}
Building upon our work, a controlled lab study looking into the efficacy of our proposed solutions, such as the consent nutrition label, applying consent mechanisms from other domains, and our suggestions for better purpose names and descriptions can yield insights into what the future of informed consent could look like. We do not measure how perceptions impact user behaviours, so a follow-up study investigating how perceptions may differ from behaviours could bring important insights. Additionally, since our study only looked at English privacy notices, future work should expand upon this by studying privacy notices in other languages to suggest improvements for non-English banners. 

\section{Conclusion}

Through semi-structured interviews with 23 UK and Ireland-based participants, we studied user perceptions of data collection purposes, which form the basis of what users are consenting to share their data for. We investigated how six common purposes (\textit{Strictly Necessary / Essential / Required, Statistics and Analytics, Performance and Functionality, Advertising, Personalized Advertising,} and \textit{Personalized Content}) were perceived by users, and identified elements of an effective purpose name and description.

Our results suggest that most purpose descriptions were not informative enough, according to participants. Descriptions do not often tell users the specifics about an organization's data handling procedures, such as how long their data is retained, nor outline the data deletion process. Overall, participants wanted descriptions to provide more transparency and reassurance. 

For purpose names, some names were preferred over others because they were perceived to be transparent or easy to understand for participants. It was common for participants to get some purposes confused with each other, or believe all purposes were covertly being used for advertising purposes. From our findings, we provide suggestions for how purpose names and descriptions can be improved, and envision a future of informed consent that is more user-centred.

{acmart}

\bibliographystyle{ACM-Reference-Format}
\bibliography{sample-base}

\end{document}